\documentclass[prl, twocolumn,superscriptaddress,nobibnotes]{revtex4-2}
\usepackage{graphicx}
\usepackage{amssymb}
\usepackage{amsmath}
\usepackage{bm}
\usepackage{color}
\usepackage{float}
\usepackage{setspace}
\usepackage{physics}
\usepackage{subfigure}
\usepackage{upgreek}

\large

\begin{document}

\title{Optically and remotely controlling localization of \\ exciton polariton condensates in a potential lattice}

\author{Qiang Ai}
\affiliation{Department of Physics, School of Science, Tianjin University, Tianjin 300072, China} 

\author{Jan Wingenbach}
\affiliation{Department of Physics and Center for Optoelectronics and Photonics Paderborn (CeOPP), Universit\"{a}t Paderborn, Warburger Strasse 100, 33098 Paderborn, Germany}
\affiliation{Institute for Photonic Quantum Systems (PhoQS),
Paderborn University, Warburger Straße 100, 33098 Paderborn, Germany}

\author{Xinmiao Yang}
\affiliation{Department of Physics, School of Science, Tianjin University, Tianjin 300072, China}

\author{Jing Wei}
\affiliation{Department of Physics, School of Science, Tianjin University, Tianjin 300072, China}


\author{Zaharias Hatzopoulos}

\affiliation{Institute of Electronic Structure and Laser (IESL), Foundation for Research and
Technology-Hellas (FORTH), Heraklion 71110, Greece 
}

\author{Pavlos G. Savvidis}
\affiliation{Key Laboratory for Quantum Materials of Zhejiang Province, Department of Physics, Westlake University,  Hangzhou, Zhejiang 310024, China}

\affiliation{Institute of Natural Sciences, Westlake Institute for Advanced Study, Hangzhou, Zhejiang 310024, China
}

\affiliation{Institute of Electronic Structure and Laser (IESL), Foundation for Research and
Technology-Hellas (FORTH), Heraklion 71110, Greece 
}

\author{Stefan Schumacher}
\affiliation{Department of Physics and Center for Optoelectronics and Photonics Paderborn (CeOPP), Universit\"{a}t Paderborn, Warburger Strasse 100, 33098 Paderborn, Germany}
\affiliation{Institute for Photonic Quantum Systems (PhoQS),
Paderborn University, Warburger Straße 100, 33098 Paderborn, Germany}%
\affiliation{Wyant College of Optical Sciences, University of Arizona, Tucson, AZ 85721, USA}

\author{Xuekai Ma}
\affiliation{Department of Physics and Center for Optoelectronics and Photonics Paderborn (CeOPP), Universit\"{a}t Paderborn, Warburger Strasse 100, 33098 Paderborn, Germany}

\author{Tingge Gao}
\thanks{Authors to whom correspondence should be addressed: tinggegao@tju.edu.cn}

\affiliation{Department of Physics, School of Science, Tianjin University, Tianjin 300072, China}

\begin{abstract}
Exciton polaritons are inherently tunable systems with adjustable potential landscape. In this work we show that exciton polariton condensates can be selectively localized in a fixed optically induced periodic lattice with uniform potential depth, by judiciously controlling a second focused pump of very small size away from the lattice chain. Specifically, the localized polariton condensate can be tuned among different potential traps by adjusting the relative distance between the small pump spot and the potential lattice. The adjustment of the excitation position of the smaller pump spot and its combination with the fixed larger pump spot for the potential creation induce the mode selection determined by gain profile, group velocity, and potential distribution within the system. The localization of the exciton polariton condensate and its control are independent of the orientation of the potential lattice, thus, even in slightly disordered system, one can selectively excite such localized polariton condensates. Our results illuminate a path to the remote manipulation of exciton polariton bosonic condensates in fixed integrated photonic chips and circuits.
\end{abstract}

\maketitle


To localize particular modes on demand is very challenging in many physical systems. It can be realized by precisely manipulating the related system parameters. By virtue of the nonlinearity induced macroscopic self-trapping, cold atom condensates can be localized in coupled potential wells or 1D/2D optical lattices \cite{PhysRevLett.95.010402, PhysRevLett.94.020403}. On the other hand, bound state in the continuum \cite{kodigala2017lasing} in photonic crystals can localize the photons with nearly infinite lifetime. Inherently non-Hermitian or PT symmetric photonic systems \cite{moiseyev2011non, konotop2016nonlinear, bender1998real} allows manipulation of the transportation behavior of photons by judiciously designing the gain and loss distribution. For example, localization of lasing modes between gain and gain-free regions is predicted in asymmetric pumped devices \cite{2011Unconventional}. The optical mode profiles can be localized in a single resonator or waveguide due to the particular distribution of the supermodes near the non-Hermitian degeneracy point (exceptional point) \cite{peng2014parity, feng2013experimental, 2014Parity}.

\begin{figure}
		\centering
		\includegraphics[width=0.85\linewidth]{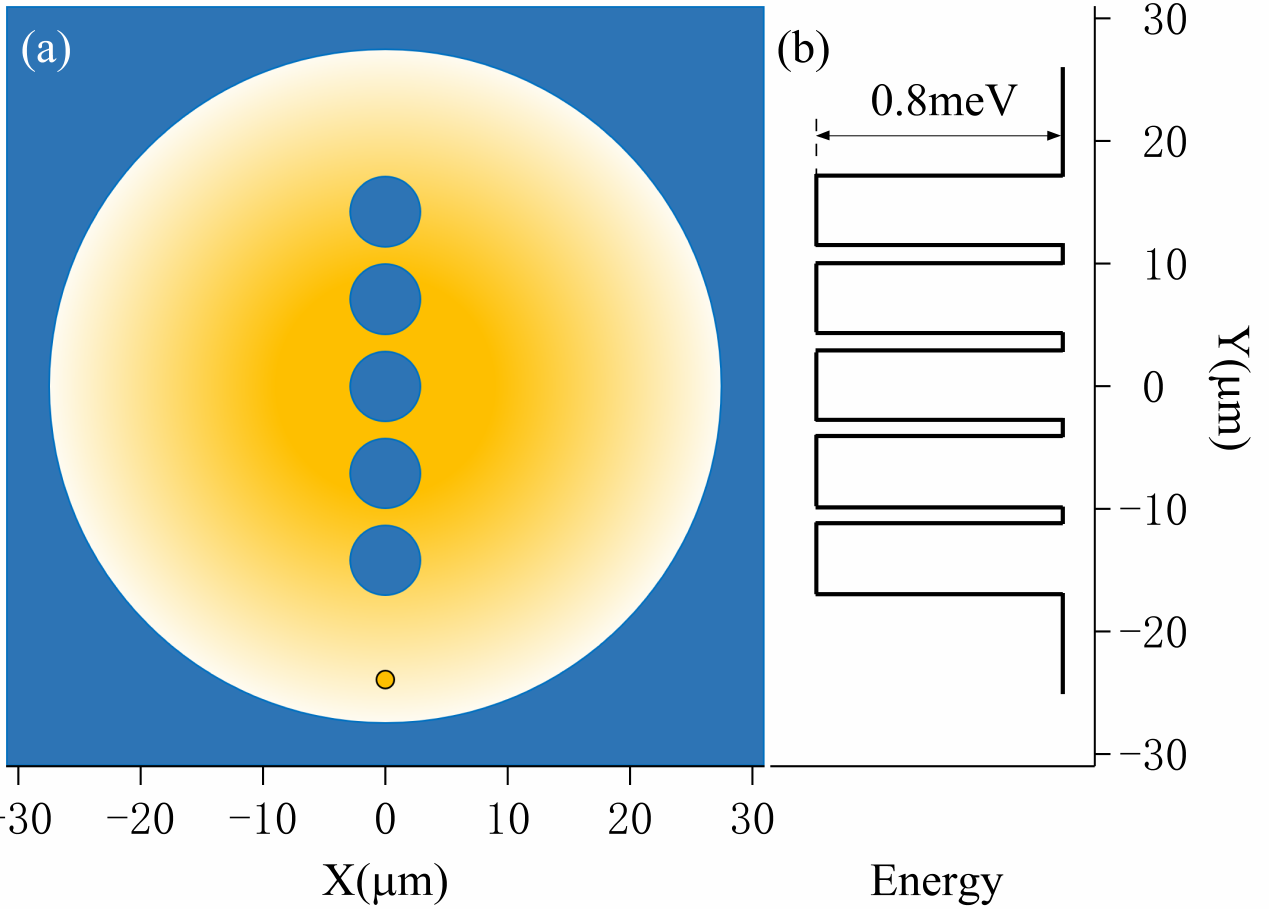}
		\caption{\textbf{Schematic diagram of the pumping configurations.} (a) The dark blue background represents the surface of the GaAs microcavity, the yellow circle is the large pumping spot with five potential traps generated by DMDs, and the orange dot is the small pumping spot. (b) The potential distribution of the 1D potential lattices with the depth of around 0.8 meV. }
	\end{figure}


As an optically controllable non-Hermitian hybrid light-matter excitation, an exciton polariton is a bosonic quasiparticle formed due to the strong coupling between the exciton and photon mode in a microcavity~\cite{kavokin2017microcavities,deng2010exciton}. It can undergo similar Bose-Einstein condensation process as the cold bosonic atoms \cite{2006Bose, balili2007bose}. Exciton polariton condensates can be localized by potential traps formed by disorder \cite{lagoudakis2008quantized,sedov2021circular}, fabricated mesa \cite{galbiati2012polariton}, optical pumping configurations \cite{gao2018chiral,tosi2012sculpting,dall2014creation,ma2020realization,gnusov2023quantum,panico2023onset}, nonlinearity under higher pumping density \cite{abbarchi2013macroscopic} or coupling with particular optical modes like the bound state in the continuum \cite{ardizzone2022polariton}. The non-Hermiticity of the polaritons affects the mode profile crucially through the gain and loss distribution, that can be actively controlled by the pumping schemes or by designing particular band structures \cite{gao2015observation,gao2018chiral,li2022switching, bao2023topological, kokhanchik2023non,PhysRevResearch.6.013148}. Exciton polaritons can also be localized at the edge of a zigzag lattice which is etched in a GaAs microcavity due to the topological edge modes \cite{st2017lasing}. By inserting or designing a particular defect into the periodic structure, polaritons can be localized at specific positions \cite{pernet2022gap,pickup2020synthetic}. However, the polariton condensate is localized by the intrinsic mode properties; changing the localized polariton distribution requires to design and fabricate different periodic structures. This does not allow to selectively manipulate the position of the localized polariton condensate within a fixed periodic structure, which is important for the remote and on-the-fly control of coherent light emission in integrated photonic chips or circuits.

In this work we show that exciton polariton condensates can be selectively localized among different potential traps in an optically formed 1D lattice of a GaAs microcavity by engineering the system's potential landscape using optical methods. Within the 1D lattice (cf. Fig.~1), the area without laser excitation can be seen as the potential trap where polaritons are confined due to the repulsive interaction with the exciton reservoir in the pumping area. In addition, we introduce a small pumping spot where a local blueshift is created and polaritons propagate radially away from the center of the pumping area. Depending on the specific pumping rate and detuning, polaritons can relax in energy from the high momentum states \cite{PhysRevB.85.235303,PhysRevLett.118.215301}. In our experiments, the polaritons further away from the small pump spot have larger group velocities and gain, which is measured and plotted in Figure S1(a-d) \cite{SM}. The polaritons propagate over the 1D chain lattice which provides the effective potential trap. We find that at specific positions away from the small excitation spot they can condense locally due to the competition between the polariton group velocity and the optically formed potential within the periodic structure. By changing only the position of the small pump spot against the fixed 1D potential lattice, we can realize precise remote control of the position where the polaritons condense. Being quite insensitive to the disorder in the sample, this kind of selective localization of polariton condensates can also be realized in differently orientated 1D potential lattices. Our work illuminates a way to selectively manipulate polariton condensates in fixed integrated photonic devices using optical means.

We use a similar GaAs microcavity as \cite{li2022switching} which is cooled in a cryostat at around 6 K. Two CW lasers are used to excite the microcavity with the wavelength of 760 nm. The first laser beam is reflected by a DMD (Digital Micromirror Device), then goes through a tube lens and is focused onto the plane of the microcavity with the size of around 40 $\upmu$m. By designing the parameters of the DMD, a 1D potential chain lattice is imaged onto the surface of the microcavity as shown in Figure 1. The potential chain lattice is composed of 5 potential traps with equal diameter. We checked the depth of the five potential sites by measuring their dispersion and the barrier dispersion below the threshold, as shown in Figure S1(e-j) \cite{SM} where equal depths of the five potential sites of around 0.8 meV can be clearly seen.  The second laser beam is focused onto the microcavity directly with the size of around 2 $\upmu$m (Figure 1). The position of the small pumping spot is shown in Figure 1 and can be tuned by moving the small pumping spot relative to the fixed 1D potential lattice. A mechanical chopper (duty circle: 5\%) is used to reduce possible heating effect which may induce the energy redshift.



We fix the pumping power of the small pump spot below the threshold at 70 mW (threshold: 130 mW), and the power of the large pump spot is set to be around 300 mW, which is smaller than the corresponding polariton condensate threshold (600 mW). The detuning of the excited area of the microcavity is around -9.5 meV. The distance between the small pump spot and the last potential trap in the 1D chain lattice is around 5 $\upmu$m. There is no condensation when only the large or small pump spot excites the microcavity separately (Figure 2(a, b)), because both of them are below their own condensation thresholds. Surprisingly, the polariton condensate is only observed at the second potential trap when both pump spots are applied. The first potential trap is not occupied by the polariton condensate (Figure 2(c)), which is confirmed by the energy-distance imaging along the 1D potential chain lattice shown in Figure 2(d). 


\begin{figure}
		\centering
		\includegraphics[width=0.7\linewidth]{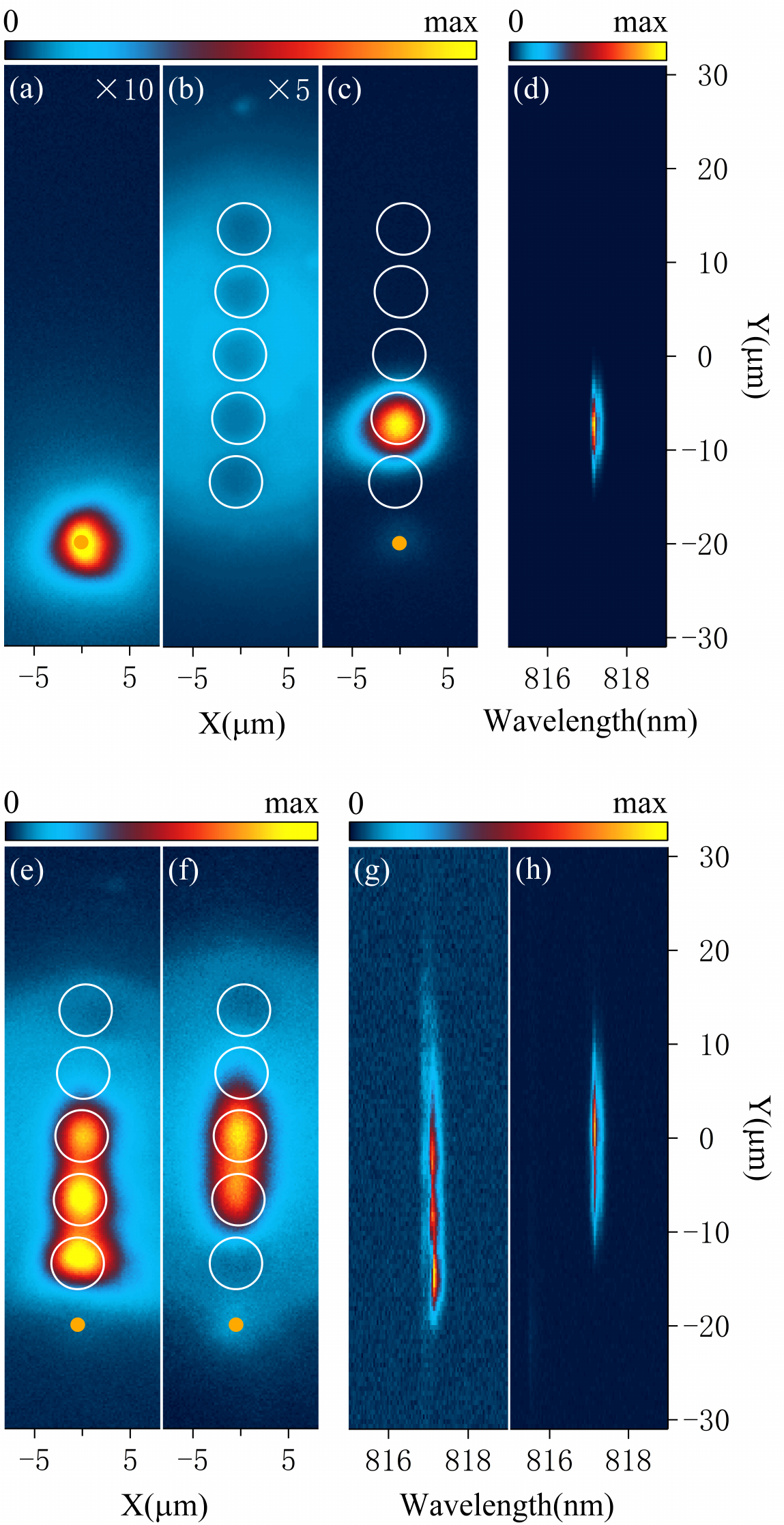}
		\caption{\textbf{Localization of the polariton condensate in the 1D lattice.} Real space imaging when only small pump spot (a, 70 mW), only 1D chain lattice (b, 300 mW), and both of them (c) excite the microcavity. (d) Energy-\textit{y} imaging along the 1D chain lattice of (c). (e, f) The polariton condensate distribution when the power of the small pump spot are 95 mW and 110 mW, respectively. (g, h) Energy-\textit{y} imaging corresponding to (e, f), respectively.}
	\end{figure}

The above studied localization of the polariton condensate is opposite to the intuition that the polariton condensate should be observed in the closest potential traps to the small pump spot together. In our experiment, the tight pumping spot creates more efficient relaxation rate and higher gain at the second potential trap (details can be found in Figure S2(d-e) \cite{SM}), thus polaritons condense locally. The numerical results shown in Figure S7 \cite{SM} can reproduce the localization of polariton condensate in the 1D chain, thanks to the real part of the potential. When the gain is larger enough in the first and third potential traps by increasing the pumping density of the small pumping spot (other parameters, such as the large pump spot and the 1D potential lattice, are fixed), we can observe that the polariton condensation occurs in these potential sites together, as shown in Figure 2(e) and Figure 2(g) (We note that the GP model is limited to simulate the experimental results in Fig. 2(e,f) in the same parameter region because of the complicated reservoir dynamics.). In Figure S2(a-c) \cite{SM} we find that multiple potential trap polariton condensation can also be observed when the tight small pump spot is replaced by a broad Gaussian-shape excitation. A larger spot size of the 1D potential chain lattice can lead to polariton condensation across the 1D potential chain, which increases the gain for all the potential traps.

The parameters of the optically formed potential distribution, including the trap dimension, spacing and depth can affect the polariton condensate localization significantly. The condensate localization at the second potential trap cannot be realized if the diameter and spacing of the potential traps are much larger or the pumping density (the potential depth) of the 1D chain is increased. Finally, the coupling strength between the potential traps also influences the localization of the polariton condensate. For instance, if the potential traps overlap partially, either condensate cannot be observed or extended polaritons appear (Figure  S4 \cite{SM}). 

\begin{figure}
	\centering
	\includegraphics[width=0.8\linewidth]{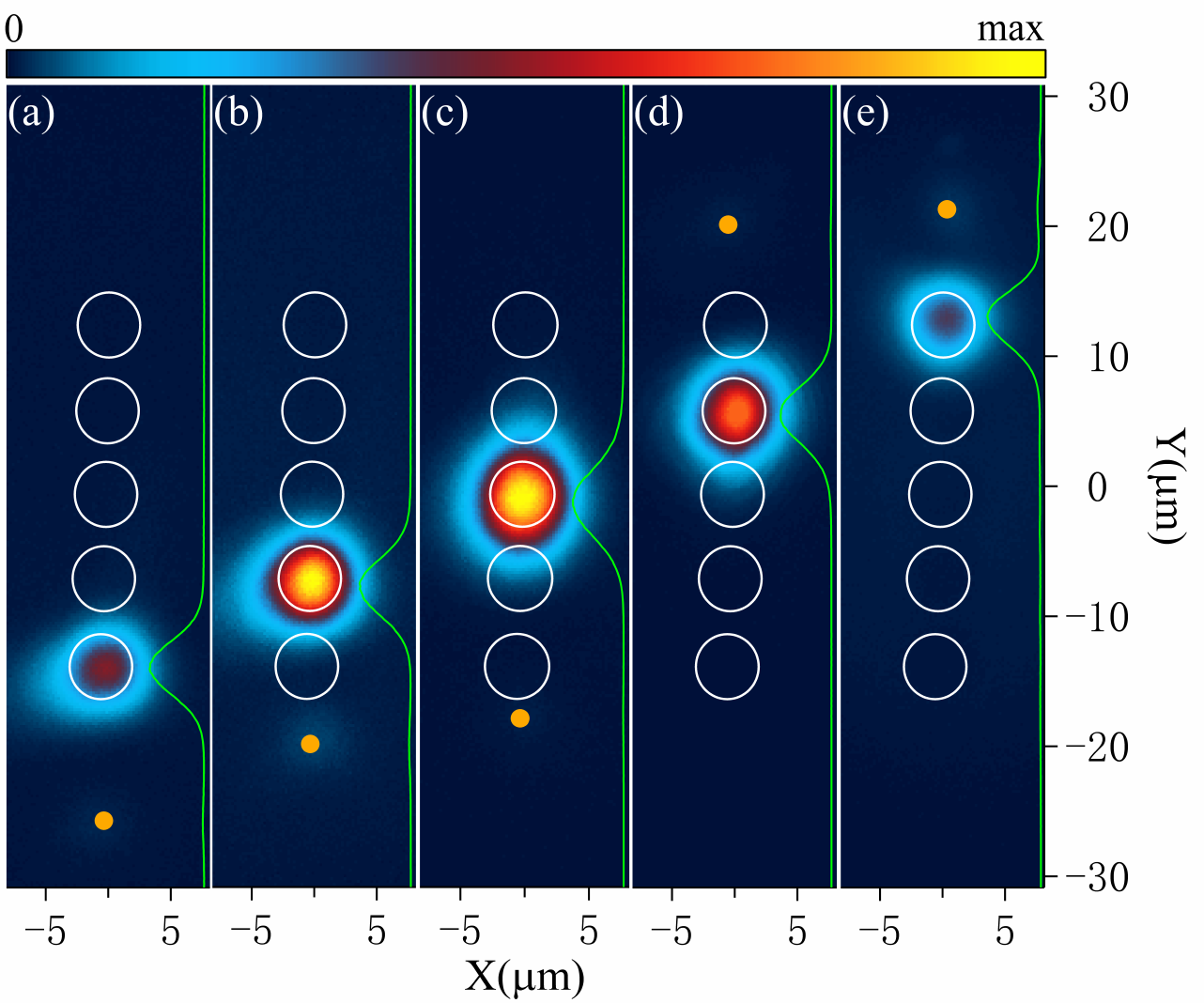}
	\caption{\textbf{Selective localization of the polariton condensate along the 1D lattice.} (a-e) Real space imaging of the polariton condensate along the 1D lattice when the position of the small pump spot is varied as indicated by the yellow dots. The inserted curves are the lineprofile along \textit{y} direction. }
\end{figure}

In the following we check the possibility that the polariton condensate at the second potential site is due to the disorder-induced trapping within the microcavity or not. To do this, the distance (period) between the potential traps is increased by around 2 $\upmu$m. The polariton condensate at the second potential trap moves accordingly (Figure S5 \cite{SM}). We also checked that these localized polaritons can be relocated to other positions of the microcavity. These can exclude the role of the disorder in the condensation process within the 1D chain lattice.


Another possible reason for the polariton condensate in the second potential trap may be due to the diffusion of hot excitons into the first potential site, which can expel the occupation of polaritons in this trap. Figure 2(e) and Figure 2(g) show that the polariton condensate occurs at the three potential traps at the same time with the same energy by increasing the pumping density of the small pumping spot. From these we can rule out the possible diffusion of hot excitons into the first potential trap resulting from the small pumping spot. The real part of the potential of the small pump spot begins to dominate the condensation process more strongly in the 1D chain lattice with further increasing the pumping density, which expels the polariton condensate further upwards, such that we do not observe the condensation in the first potential trap plotted in Figure 2(f) and 2(h). Note that the energies of the localized and extended polariton condensates are the same (Figure 2(d, g, h)), confirming the absence of a heating effect and redshift.





Selectively tuning the polariton condensate position can be realized by adjusting the distance between the small pump spot and the 1D chain lattice, which enables the polaritons to propagate over different distances, thus the single potential traps above (the third) or below (the first) can be occupied by the condensate. By moving the small pump spot downwards or upwards to keep away from or approach the chain lattice with fixing the pumping density unchanged, we observe that the polariton condensate can be localized in the first and third potential site consecutively (Figure 3(a, c)). It is worth noting that polaritons do not condense when the small pump spot is placed at the side of the second and third potential traps. The polariton condensate in the fourth and fifth potential trap can be realized by placing the small pump spot to the top of the chain lattice, as shown in Figure 3(d, e).  The lineprofile across the 1D chain lattices clearly shows the localization of the polariton condensate. We check the energy of the trapped polariton condensate in different potential sites. All the condensates share the same energy, this means there is no different potential depth distribution along the 1D lattice. 




\begin{figure}
	\centering
	\includegraphics[width=\linewidth]{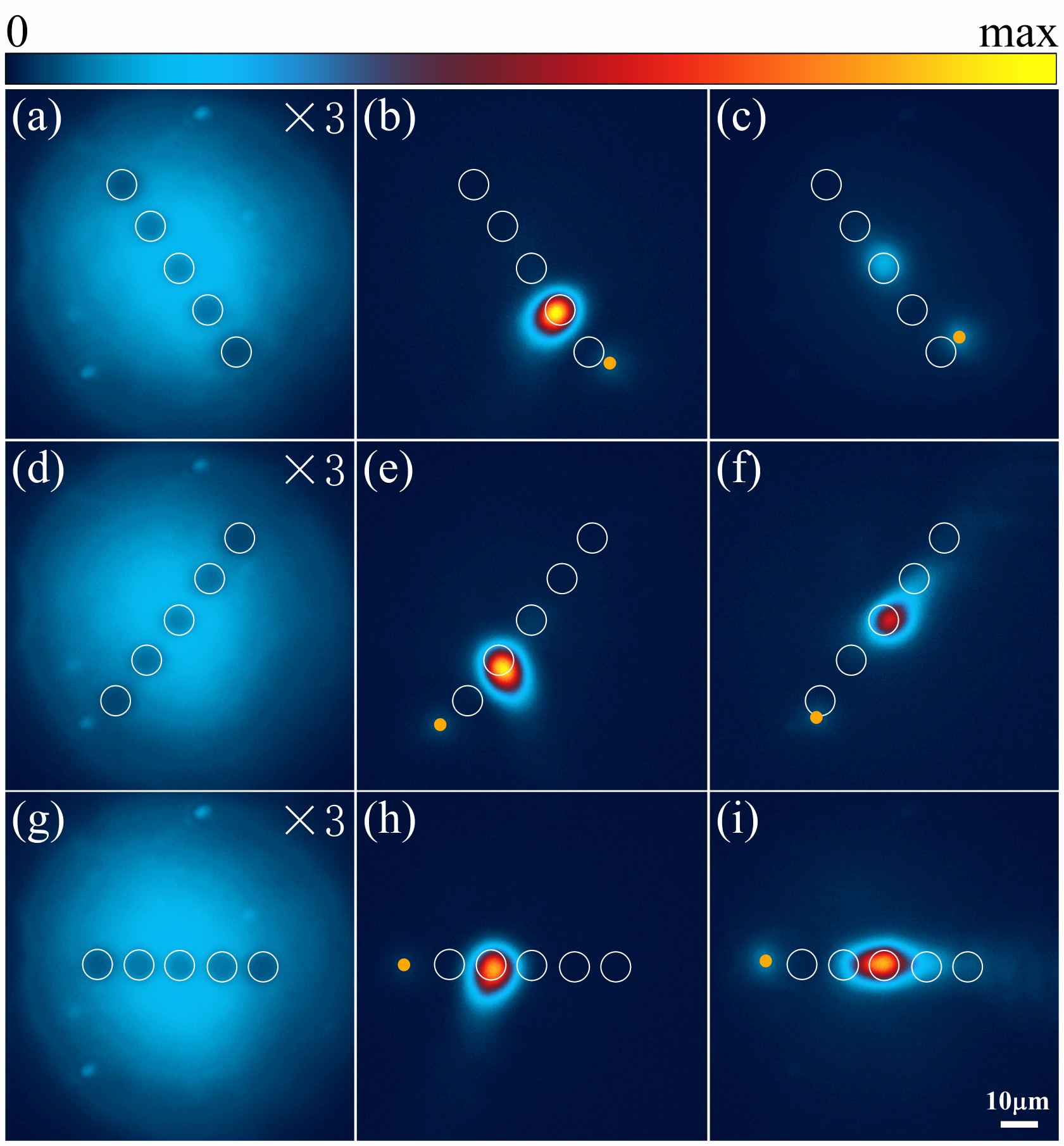}
	\caption{\textbf{Selective localization of the polariton condensate in differently oriented potential lattices.} (a, d, g) The pumping configuration along antidiagonal, diagonal and horizontal direction below the threshold. (b, c), (e, f) and (h, i) Polariton condensate localized at the second (third) potential site corresponding to (a, d, g).   }
\end{figure}

The localization of the polariton condensate in the single potential trap is not limited by the orientation of the chain lattice. In the following we re-arrange the direction of the chain lattice in antidiagonal (Figure 5(a)), diagonal (Figure 5(d)) and horizontal (Figure 5(g)) directions along the microcavity plane. In these experiments, the small pump spot is placed at the left or bottom of the last potential traps. In all these different chain lattices, polaritons can condense in the single potential traps (Figure 5(b), Figure 5(e) and Figure 5(h)). In addition, the localization of the polariton condensate in these different 1D chain lattices can be controlled by adjusting the position of the small pump spot (Figure 5(c), Figure 5(f) and Figure 5(i)). This further confirms that the disorder does not contribute to the localization of the polariton condensate. Especially in Figure 5(c) the polaritons can be localized in the middle potential trap when the small pump spot is placed at the side of the first potential site. On the other hand, polaritons do not condense in the nearest potential trap if the tight pump spot is placed at the side of the 1D chain lattice (Figure 5(c)), which shows the advantage of the remote control of polariton condensate distribution in our experiment. 


To conclude, we realized selective localization of polariton condensates in a fixed 1D optically induced potential chain lattice by tuning the position of a second small pump spot away from the 1D chain. Proving itself quite insensitive to disorder, the localization does not depend on the orientation of the 1D lattice. Our work provides an approach to remotely control the polariton condensate using simple optical methods and can find applications in future optoelectronic circuits based on exciton polariton condensates.

\begin{acknowledgments}
TG acknowledges the support from the National Natural Science Foundation of China (NSFC, No. $12174285$). The Paderborn group acknowledges the support from the Deutsche Forschungsgemeinschaft (DFG) (Grant No. 467358803 and No. 519608013) and by Paderborn Center for Parallel Computing, PC$^2$.
\end{acknowledgments}
%


\end{document}